\documentclass[aps,prl,twocolumn,showpacs]{revtex4}
\usepackage{amsmath,amsfonts,bm, color,amssymb}
\usepackage{graphicx,epsfig,overpic,hyperref}

\input epsf 

\newcommand{\Rr}{\mbox{\it R}}
\newcommand{\Sr}{\mbox{\it S}}

\newcommand{\refeq}[1]{Eq. \ref{#1}}
\newcommand{\sigm}{{\sigma}}
\newcommand{\out}{{\mathrm{ex}}}
\newcommand{\ins}{{\mathrm{in}}}

\newcommand{\bE}{{\bf E}}
\newcommand{\bP}{{\bf{P}}}

\begin{document}

\title{Nonlinear electrohydrodynamics of a viscous droplet}
\author{Paul F.  Salipante  and Petia M. Vlahovska, }
\affiliation{%
 School of Engineering, Brown University, Providence, RI 02912
}

\date{\today}

\begin{abstract}

A classic result due to G.I.Taylor is that a drop placed in a
 uniform electric field becomes a prolate or oblate spheroid, which is axisymmetrically aligned with the applied field. We report an instability and symmetry-breaking transition to  obliquely oriented, steady and unsteady shapes  in strong fields.
Our experiments reveal novel droplet behaviors such as tumbling, shape oscillations, and chaotic dynamics even under creeping flow conditions. A theoretical model, which includes anisotropy in the polarization relaxation due to drop asphericity and charge convection due to drop fluidity, elucidates the interplay of interfacial flow and charging as the source of the rich nonlinear dynamics.

\end{abstract}

\maketitle

Nonlinear phenomena such as instabilities and turbulence naturally occur in fluid dynamics because of the nonlinearity of the Navier-Stokes due to the inertial term.
 In the absence of inertia, the Stokes equations governing the fluid flow are linear and evolving boundary conditions
are the only  source of nonlinearity \cite{Blawzdziewicz:2010}.    For example, a single particle exhibits complex dynamics  
if it is deformable:    a capsule or a red blood cell in shear flow \cite{Omori:2012, Gao:2012, Skotheim:2007, Abkarian:2010, Vlahovska_swing:2011}, a drop in oscillatory strain-dominated linear flows \cite{YJ2008}, or a drop sedimenting in an electric field \cite{Homsy}.

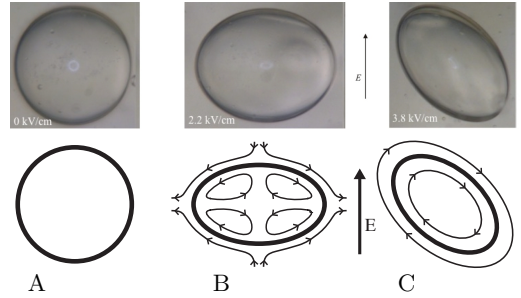
\begin{figure}[h]
  \begin{picture}(0,0)(0,0)
\put(10,-8){A}
\put(80,-8){B}
\put(150,-8){C}
\end{picture}
 \includegraphics[height=1.4in]{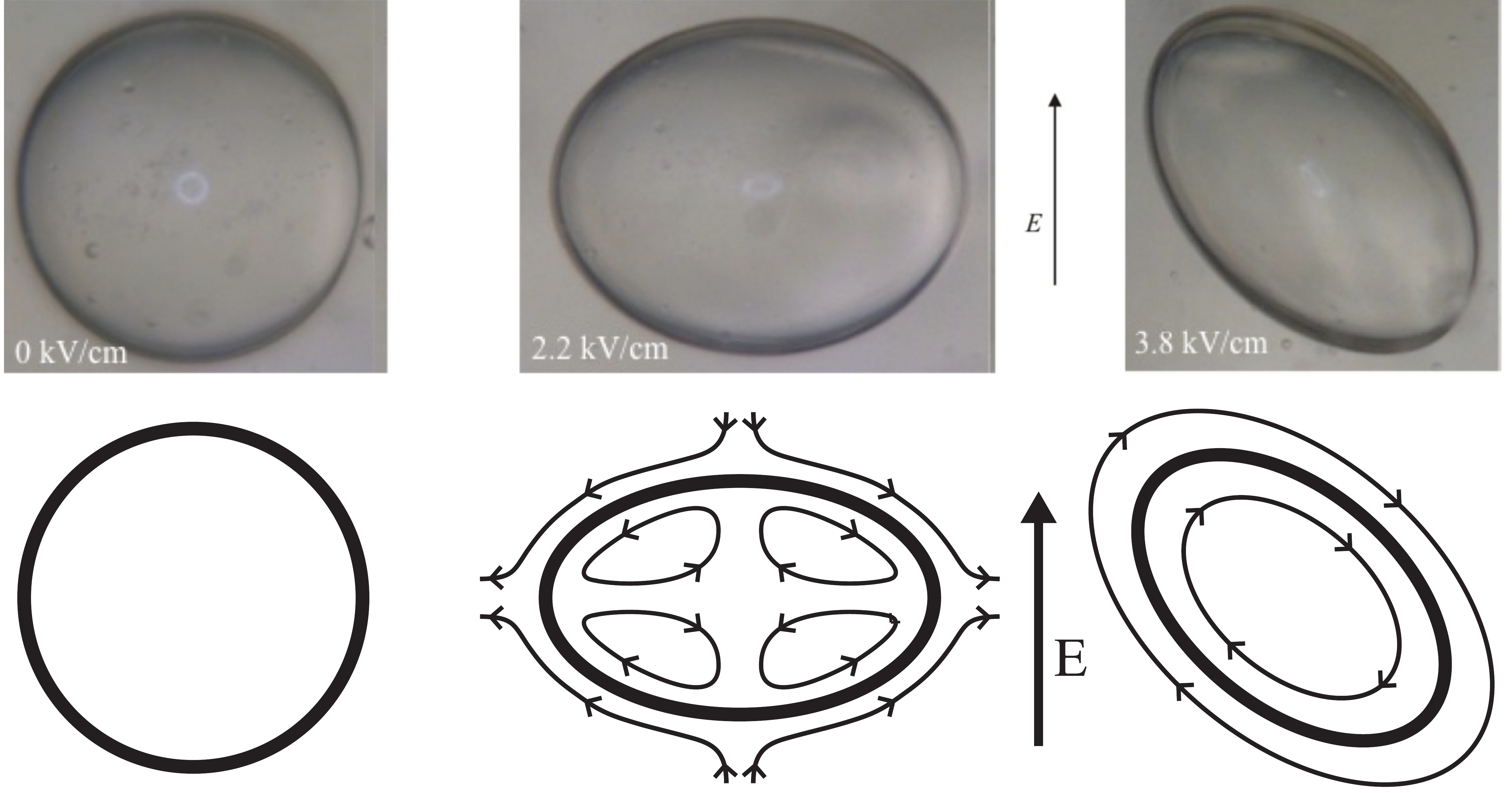}
      \caption{\footnotesize Exposure to a uniform direct current (DC) electric field with increasing strength excites a variety of drop responses \cite{Salipante-Vlahovska:2010} (see videos \cite{supplem}). A. A castor oil drop suspended in silicon oil is spherical in the absence of electric field. B. Weak fields induce axisymmetric oblate deformation. C. In strong fields, the drop is tilted with respect to the applied field direction  and the flow has a  rotational component.
      The sketches illustrate the drop shape and flow streamlines.}
       \label{figP}
  \end{figure}

If the particle is rigid, 
chaotic motions are observed with ellipsoids in  shear flow \cite{Yarin} or uniform electric fields \cite{Lemaire:2002, Cebers:2000}. In the latter case, the  nonlinear dynamics arises from anisotropy in the polarization relaxation due to asphericity of the particle shape.  A question arises  - what if the particle is not rigid, but fluid and its shape is not fixed? Would particle fluidity and deformability give rise to new, richer behaviors?
In this Letter, we explore these questions on the example of a viscous drop subjected to a uniform direct current (DC) electric field. Upon increase in the field strength, this system was found to undergo a symmetry--breaking transition from axisymmetric to linear flow  illustrated in  Figure \ref{figP}. B and C \cite{Salipante-Vlahovska:2010}, which hints upon the possibility of more complex physics. 
   We first discuss nonlinear drop electrohydrodynamics theoretically, and then we describe its experimental realization. 
 We find that depending on the fluids viscosity, the droplet may exhibit a range of intriguing dynamics such as tumbling, sustained shape oscillations, and even chaos. 

\begin{figure}[[h]
\centerline{\includegraphics[width=3.25in]{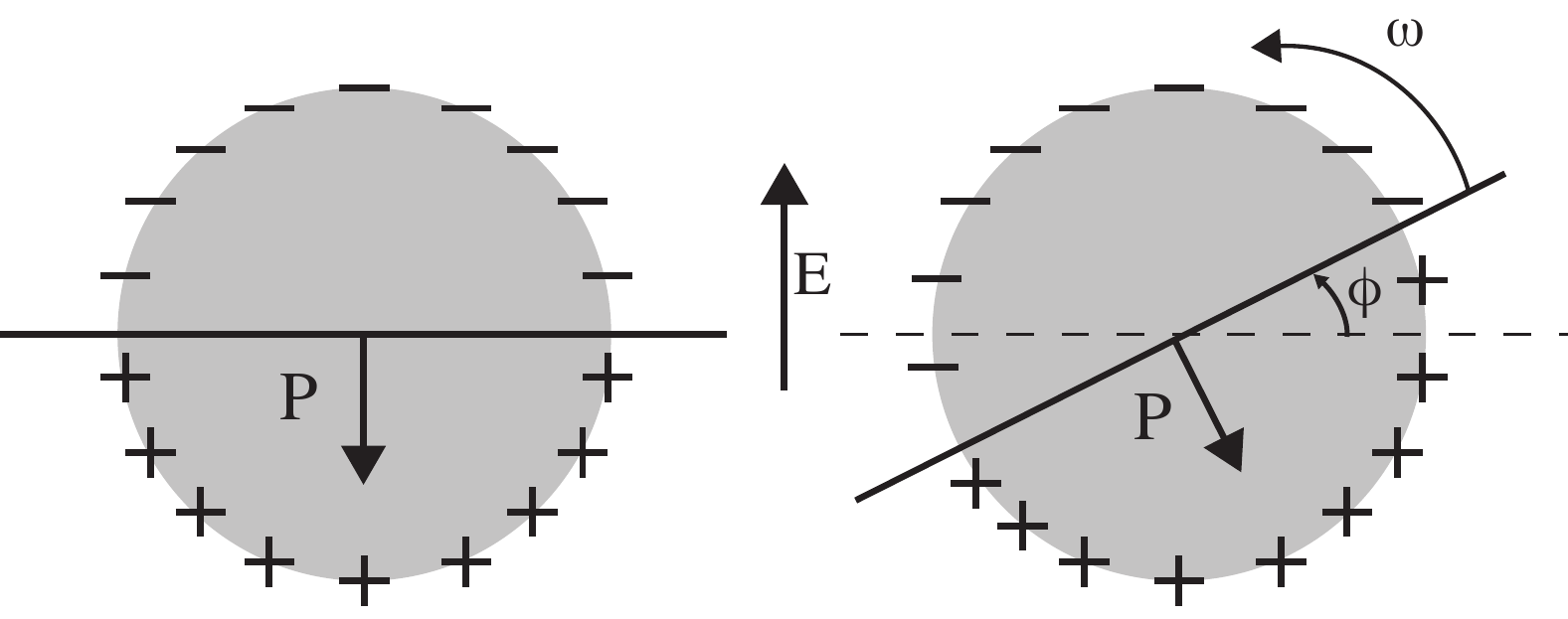}}
\caption{\footnotesize Charge distribution and induced dipole ${\bf{P}}$ 
 for a sphere with $RS<1$.  Above a critical field strength $E_0>E_Q$ , where $E_0=|\bE|$ and $E_Q$ is given by \refeq{tMW},  constant rotation around an axis perpendicular to the electric field  is induced by the misaligned dipole moment of the particle (right). The rotation can be either clock- or counter-clockwise.  
}
\label{Quinckefigure}
\end{figure}

{\it{Theoretical model for drop polarization and shape evolution.}}
When placed in an electric field, a  particle  
polarizes because free charges carried by conduction  accumulate at boundaries that separate media with different electric properties. This is illustrated in Figure \ref{Quinckefigure} on the example of a sphere in a uniform electric field. The magnitude and orientation of the induced dipole depend on  the mismatch of electric properties  between the particle ($``\ins "$) and  the suspending fluid ($``\out "$)
\begin{equation}
\Rr=\frac{\sigm_\ins}{\sigm_\out}\,,\quad \Sr=\frac{\epsilon_\out}{\epsilon_\ins}\,,
\label{ParameterRatios}
\end{equation}
where $\sigm$ and $\epsilon$ denote conductivity and dielectric constant, respectively.
The product of $R$ and $S$ compares 
the charge relaxation times of the media \cite{Melcher-Taylor:1969, Saville:1997}
\begin{equation}
\label{tc}
RS=\frac{\tau_{c,\out}}{\tau_{c,\ins}}\,,\quad\mbox{where} \quad \tau_{c,\ins}=\frac{\epsilon_{\ins}}{\sigma_{\ins}}\, \quad \tau_{c,\out}=\frac{\epsilon_{\out}}{\sigma_{\out}}\,.
\end{equation} 
If $RS<1$ ($\tau_{c,\ins}>\tau_{c,\out}$), the conduction response of the exterior fluid is faster than the particle material.  
As a result,
the induced dipole is oriented opposite to the applied electric field direction.
This configuration is unfavorable and  becomes unstable above a critical strength of the electric field \cite{JonesTB:1984,Turcu:1987, Lemaire:2002}.  A perturbation in the dipole alignment  gives rise to  a torque, which drives physical rotation of the sphere.  The induced surface-charge distribution rotates with the particle, but at the same time the exterior fluid  recharges the interface. The balance between charge convection by rotation and supply by conduction from the bulk results in an oblique dipole orientation shown in Figure \ref{Quinckefigure}.(b).   
The rotation rate $\omega$ is determined from the 
balance of electric and viscous torques acting on the particle, 
    $\bP\times\bE={\bf {A}} \cdot{\bm \omega}$, where ${\bf{A}}$ is the friction matrix. 
 
  The spontaneous spinning of a rigid sphere in a uniform DC electric field has been known for over a century, first attributed to the work of Quincke in 1896. In this case,
the friction matrix  is diagonal, and 
a straightforward calculation \cite{Turcu:1987, JonesTB:1984, JonesTB}, assuming instantaneous polarization, yields three possible solutions:
no rotation, $\omega=0$,  and 
 \begin{equation}
\label{quinckeW}
\begin{split}
 {\omega=\pm \frac{1}{\tau_{mw}}\sqrt{\frac{E_0^2}{E_Q^2}-1}\,,}
 \end{split}
\end{equation}
where the $\pm$ sign reflects the two possible directions of rotation and 
 \begin{equation}
\label{tMW}
{\tau_{mw}=\frac{\epsilon_{\ins}+2\epsilon_{\out}}{\sigma_{\ins}+2\sigma_{\out}} \quad \mbox{and}\quad E_Q^2=\frac{2\sigm_\out \mu_\out \left(R+2\right)^2}{3\epsilon_\out \epsilon_\ins (1-RS)}\,.}
\end{equation}
$\tau_{mw}$, the Maxwell-Wagner polarization time, is the characteristic time-scale for polarization relaxation. 
The  dipole ``tilt''  is steady; the angle between the dipole and the electric field 
is $\phi=\arctan\left[(\tau_{mw}\omega)^{-1}\right]$.
 \refeq{quinckeW} shows that rotation is possible only if the electric field exceeds a critical value given by $E_Q$. Hence, if $E_0\le E_Q$, the sphere and the suspending fluid are motionless. If $E_0>E_Q$, the  sphere rotates and drags the fluid in motion; the resulting flow is purely rotational.

If  polarization relaxation, i.e., non-instantaneous charging of the interface described by $\partial_t \bP={\bm\omega}\times \bP-\tau_{mw}^{-1}\left(\bP-\bP_{eq}\right)$,   is included in the analysis, the polarization evolution equation and the torque balance map onto the Lorenz chaos equations \cite{Lemaire:2002, Lemaire:2005}. A  second bifurcation occurs in stronger fields and  the sphere exhibits chaotic reversal of the rotation direction.

Unlike solid particles, drops are fluid and have a free boundary. The electric stress deforms the drop and, as a result, the friction matrix and polarization relaxation become anisotropic and dependent on the drop orientation relative to the applied electric field. Moreover, the interface does not move as a rigid body  and its velocity differs from that the Quincke rotation. The modified torque balance (assuming for simplicity that drop shape remains an axisymmetric oblate spheroid with fixed aspect ratio once rotation is initiated) is
\begin{figure}[h]
\centerline{\includegraphics[width=2.0in]{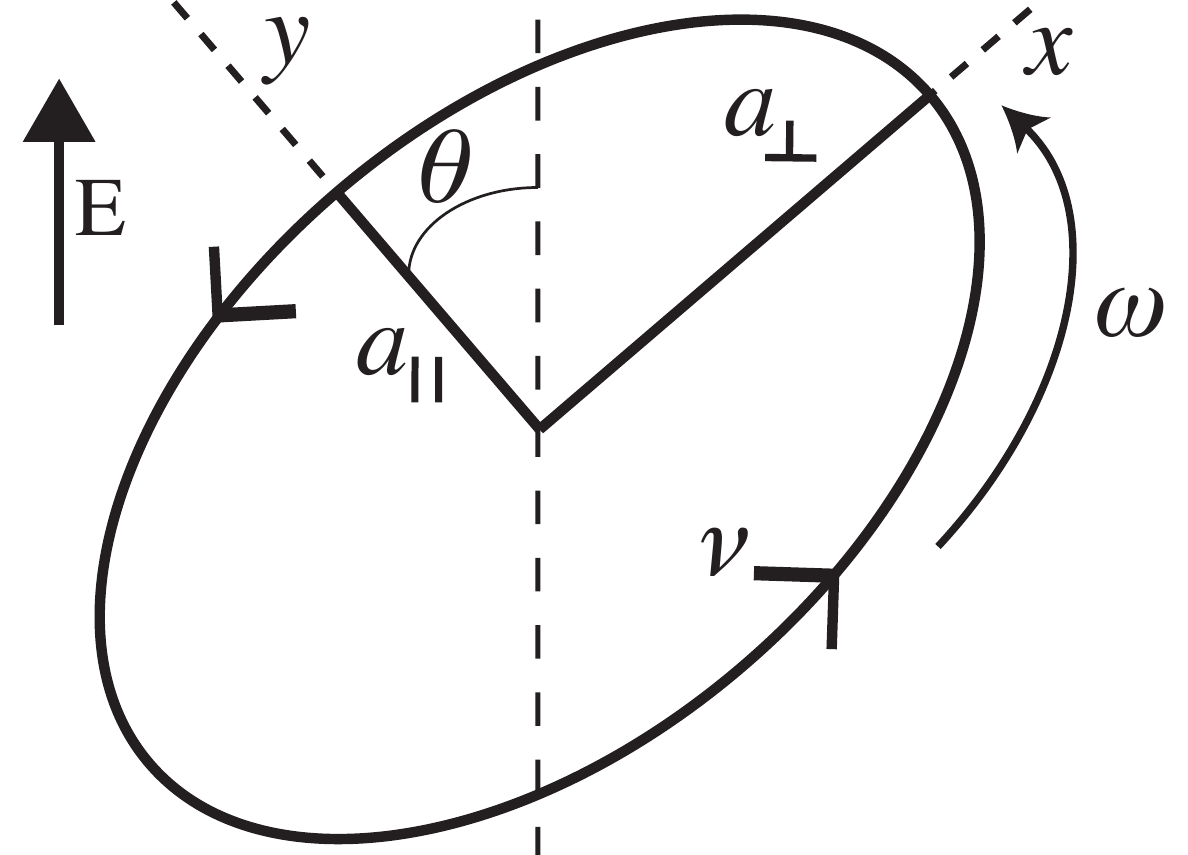}}
\caption{\footnotesize Sketch of the model for drop electrorotation. Drop asphericity is characterized by the aspect ratio $\beta$=${a_\parallel}/{a_\perp}$. The rigid body rotation is related to the tilt angle of the deformed shape, $\theta$, by $\omega={d \theta}/{dt}$. In the co-rotating frame the interface velocity is  $\nu(-y \beta, x/{\beta})$.
}
\label{fig3a}
\end{figure}
\begin{equation}
\alpha_{\perp}\left( \frac{d \theta}{dt}+ \frac{2 \beta}{1+\beta^2} \nu\right)=P_{ \parallel} E_{\perp}-P_{\perp} E_{ \parallel}+(\chi_{\parallel}^{\infty}-\chi_{\perp}^{\infty})E_{\perp}E_{ \parallel}
\label{Tbalance}
\end{equation}
where $\alpha_{\perp}$ is the friction coefficient, $\chi^\infty$ is the high-frequency susceptibility, and $\parallel$, $\perp$ denote components parallel and perpendicular to the axis of symmetry, see Figure \ref{fig3a}.  Drop fluidity is accounted by interfacial velocity with  frequency $\nu$. The latter is determined from the balance of electrical energy  input and  viscous dissipation, in a similar fashion to the analysis of drop rotation  in a magnetic field \cite{Lebedev2003}, or red blood cell tank-treading in shear flow  \cite{Keller-Skalak:1982}.
 The polarization relaxation equations in a coordinate system rotating with $\omega$ are \begin{subequations}
\begin{align}
		&\frac{\partial P_{\parallel}}{\partial t}=-\nu P_{\perp}\beta-\frac{1}{\tau_{\parallel}}[P_{\parallel}-(\chi_{\parallel}^0-\chi_{\parallel}^{\infty})E_{\parallel}] \, , \\
		&\frac{\partial P_{\perp}}{\partial t}=\frac{\nu P_{\parallel}}{\beta}-\frac{1}{\tau_{\perp}}[P_{\perp}-(\chi_{\perp}^0-\chi_{\perp}^{\infty})E_{\perp}] \,,
\end{align}
\label{polar}
\end{subequations}
where $\tau_{\parallel,\perp}$ and $\chi^0_{\parallel,\perp}$ are directional Maxwell-Wagner relaxation timescales and low-frequency susceptibility respectively.  If $\nu=0$, \refeq{Tbalance} and  \refeq{polar} reduce to the equations of motion for a rigid ellipsoid \cite{Cebers:2000}, and predict three types of behavior: alignment of the symmetry axis with the electric field, oscillations around the field direction (``swinging"), and  continuous flipping (``tumbling").  The additional torque associated with the interface fluidity (characterized by  $\nu$) modifies these behaviors to: steady axisymmetric orientation (Taylor regime), steady tilted orientation, swinging around a non-zero tilt angle with respect to the electric field, and tumbling. Moreover, chaotic switching between the swinging and tumbling states is also found. The phase diagram resulting from the numerical solution of \refeq{Tbalance} and  \refeq{polar} is shown in Figure \ref{fig4}. In all cases except for the chaotic behavior, the motion of the dipole mirrors the motion of the ellipsoid. The model predicts that chaos is observed with high viscosity drops  (drop viscosity at least  four times greater than the suspending medium).  In this case, the chaotic motion stems from the lack of synchronization between ellipsoid and dipole orientations. 

\begin{figure}[[h]
\centerline{\includegraphics[width=3in]{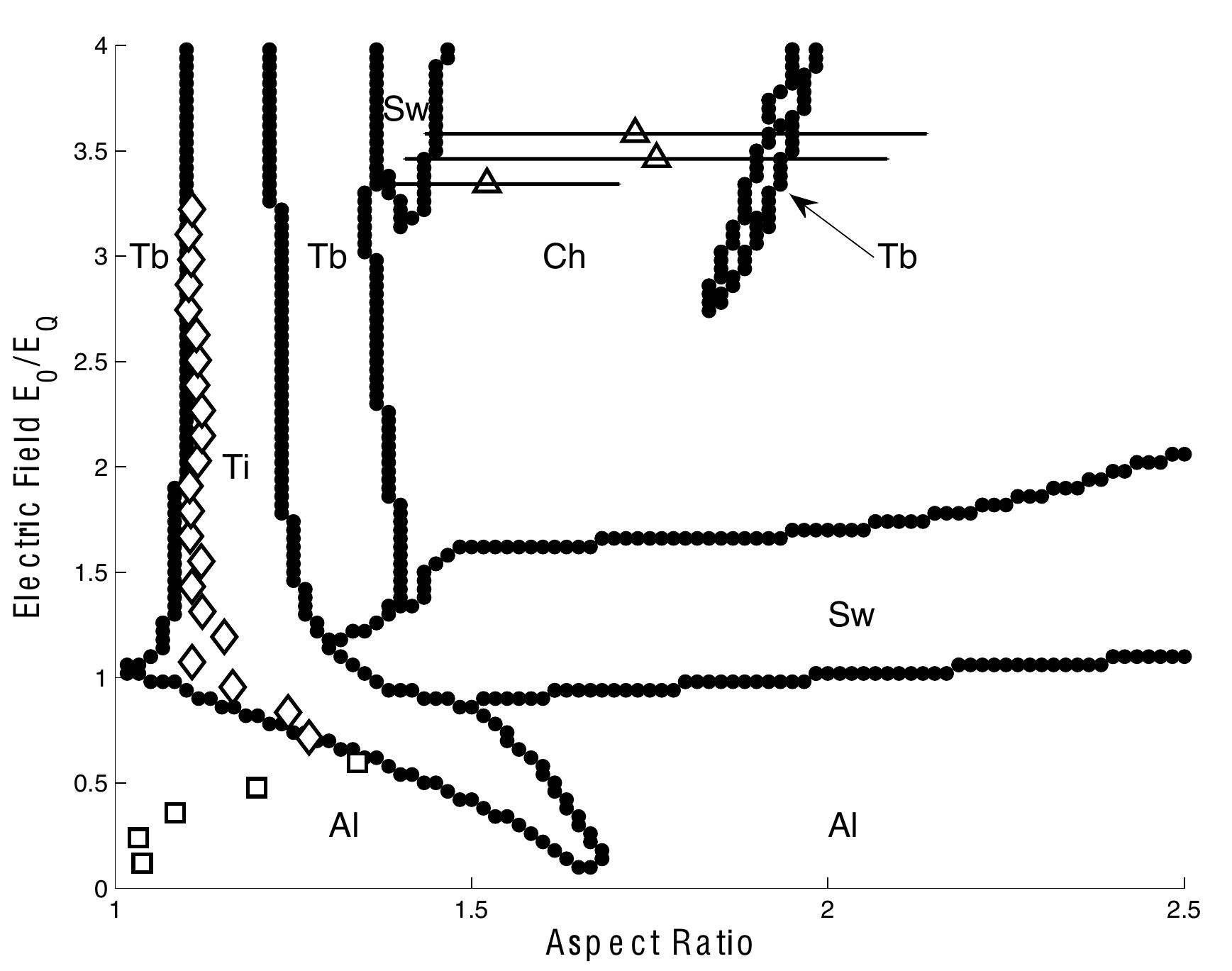}}
\caption{\footnotesize Phase diagram for a fluid ellipsoid with viscosity ratio $\lambda\equiv\mu_\ins/\mu_\out=14$. The lines represent the boundaries between various behaviors computed from the model. The symbols correspond to experimental data: $\square$ - Taylor (steady axisymmetrically oriented ellipsoid), $\diamond$ - steady  tilted ellipsoid, $\triangle$ - chaotic tumbling. The lines crossing the symbols reflect the variations in the aspect ratio of the drop during the rotation.
}
\label{fig4}
\end{figure}

\begin{figure}[h!]
\centerline{\includegraphics[width=3.5in]{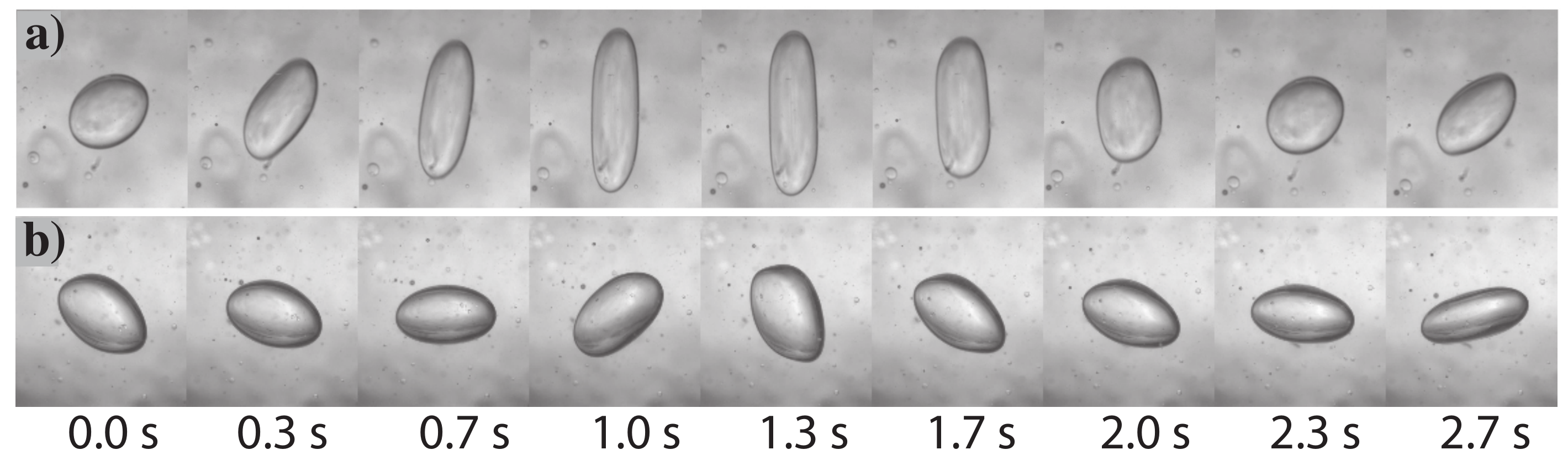}}
\caption{\footnotesize Examples of oscillatory  and tumbling drop behavior.  a) $\lambda$=1, $E_0$=9.9kV/cm, a=1.8mm. 
b) $\lambda$=14, $E_0$=9.7kV/cm, a=3.0mm
}
\label{fig5}
\end{figure}

Of course in reality drop shape does not remain an ellipsoid with fixed aspect ratio, and the experiments show pronounced shape oscillations in the case of low-viscosity drops (see Figure \ref{fig5}.a). This is due to the fact that the rotation period is comparable with the surface tension relaxation time $a\mu_\out/\gamma$ ($\gamma$ being the interfacial tension, $\mu_\out$ the suspending fluid medium,  and $a$ the initial drop radius). However, in this case the main axis is found to oscillate around a defined tilted angle and hence this behavior qualitatively corresponds to the swinging mode.  High-viscosity drops remain nearly undeformed throughout one period of rotation and tumble like rigid ellipsoids (see Figure \ref{fig5}.b)

Our ``fluid ellipsoid" model for drop electrorotation identifies drop viscosity as a key control parameter for drop behavior. Next we experimentally test the model and show that it qualitatively captures the variety of drop responses.

{\it{Experiment:}}  Weakly conducting fluids are used for the drop (silicone oil) and continuous phase (castor oil). This experimental system is characterized by  conductivity ratio $R=0.03$ and permittivity ratio $S=1.8$. To investigate the effect of drop viscosity, the viscosity ratio $\lambda=\mu_\ins/\mu_\out$ is varied in the range 1 to 14. The electric field is increased in small steps of about $0.1 E_Q$ and the system is allowed to equilibrate at each step to avoid spurious transients. For example, it is possible to excite drop tumbling upon a sudden increase of the field strength but the drop  eventually settles into steady tilted state.   More details about the experiment can be found in \cite{Salipante-Vlahovska:2010}.

Typical drop behaviors are illustrated in Figure \ref{fig6}. The unsteady dynamics of high viscosity drops ($\lambda=14$) is chaotic tumbling, while low viscosity drops ($\lambda=1$) undergo steady shape oscillations. Intermediate viscosity drops ($\lambda=4$) exhibit both behaviors, namely a cycle consisting of shape oscillations with increasing amplitude followed by several tumbles. These behaviors are better seen in the insets of Figure \ref{fig6}, where the angle between the drop major axis and the applied field direction is plotted.

For high viscosity drops $\lambda=14$, unsteady behavior is observed at aspect ratio above $\beta> 1.2$ in both the model and experiment.  Experimentally, we observe a perturbation to the oblate drop shape to slowly grow until reaching a chaotic tumbling state, in which the direction of rotation reverses chaotically. The transitions between the Taylor, tilted and  tumbling states of a drop are  given by lines in Figure \ref{fig6}.a. 
The largest stretching  occurs while the short axis of the oblate  drop is aligned with the field (see Figure \ref{fig5}.b at 0.7s and 2.37s). At this point, the drop can either break up or continue to  rotate.

 Low viscosity drops ( $\lambda=1$) exhibit steady oscillations between a prolate  and a tilted oblate shapes, see Figure \ref{fig5}.a. These shape oscillations can be explained by considering the dynamics of the induced dipole and shape. 
The lower fluid viscosity allows the drop to be easily deformed by the electric field acting on the induced surface charge. As the interface charges, the drop is squeezed into an oblate shape. The accumulated surface charge rotates due to the electric torque until the induced dipole is nearly aligned with the field. In this dipole configuration, the drop is pulled into a prolate shape until the surface charge relaxes.  The cycle repeats by the interface recharging and deforming into an oblate shape. 
At fields strengths well above the transition to unsteady behavior, the amplitude and frequency of oscillation increase until the drop stretches into a highly deformed prolate "dumbbell" shape, which then either undergo breakup or collapses back into itself, disrupting its rotational velocity. After collapsing, the oscillations will begin to build again in the same direction as before. 

\begin{figure}[h!]
\label{fig6}
\begin{overpic}[width=3.2in]{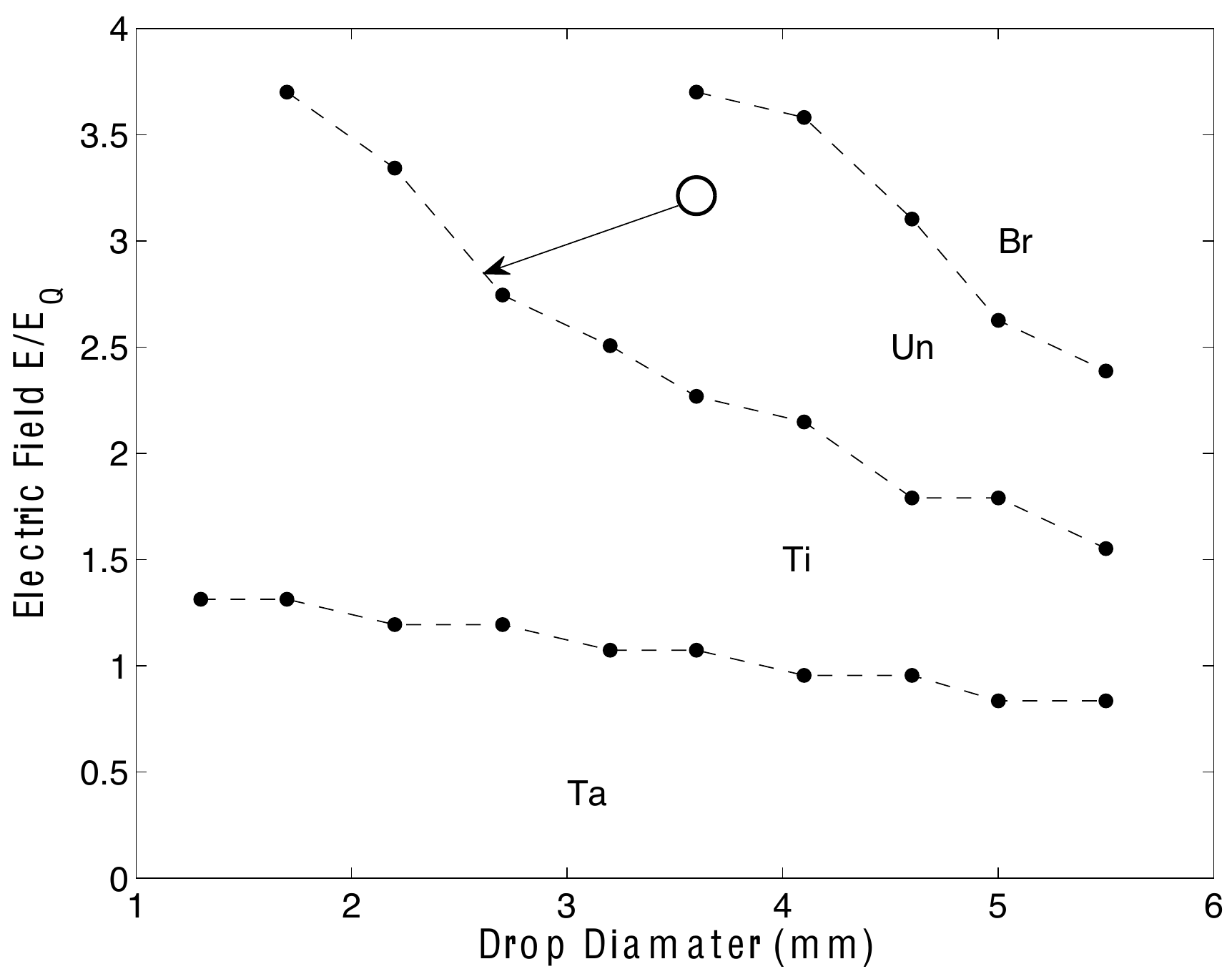}
  \put(10,31){\includegraphics[width=1.1in]%
      {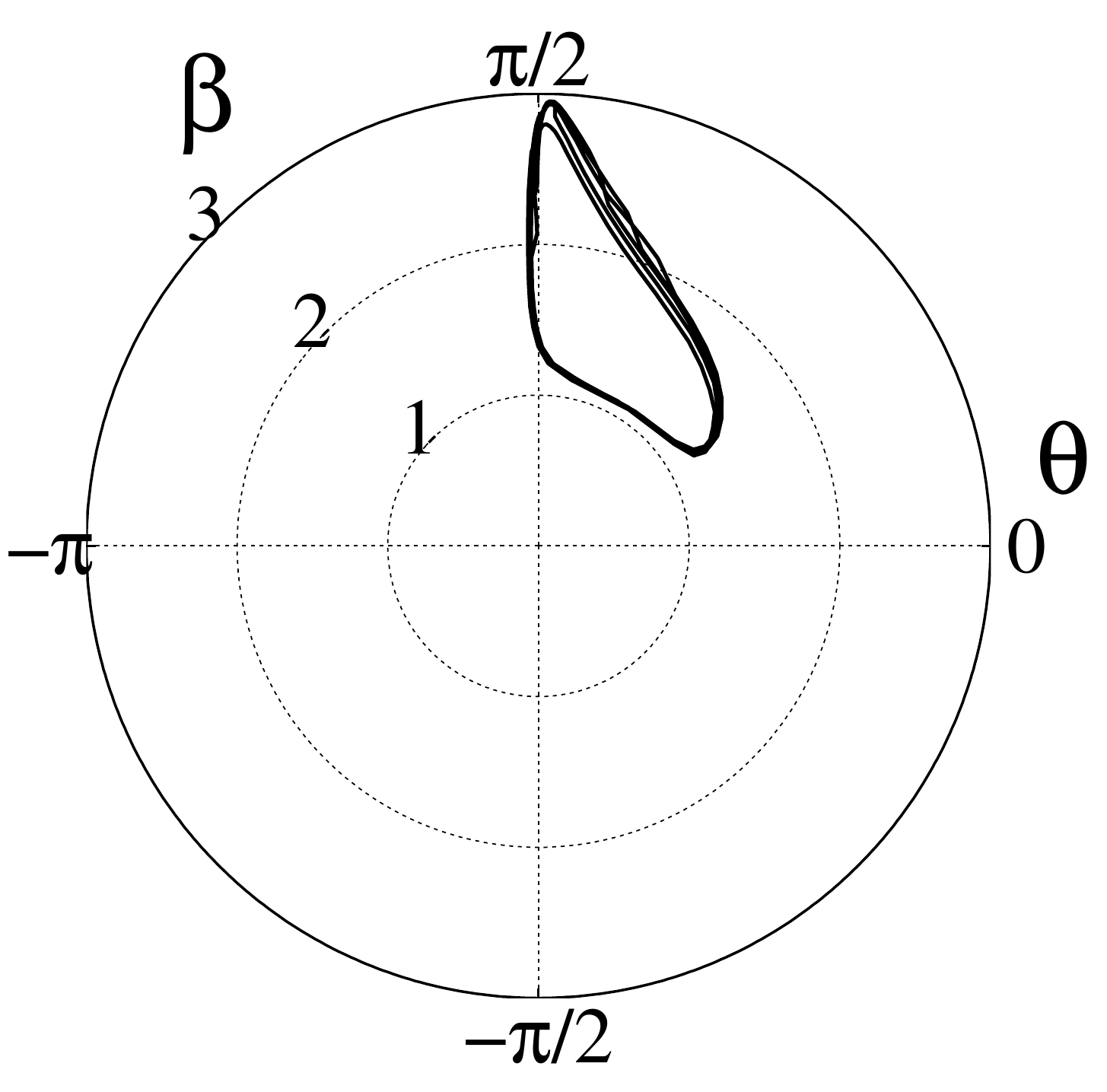}}
\end{overpic}

\begin{overpic}[width=3.15in]{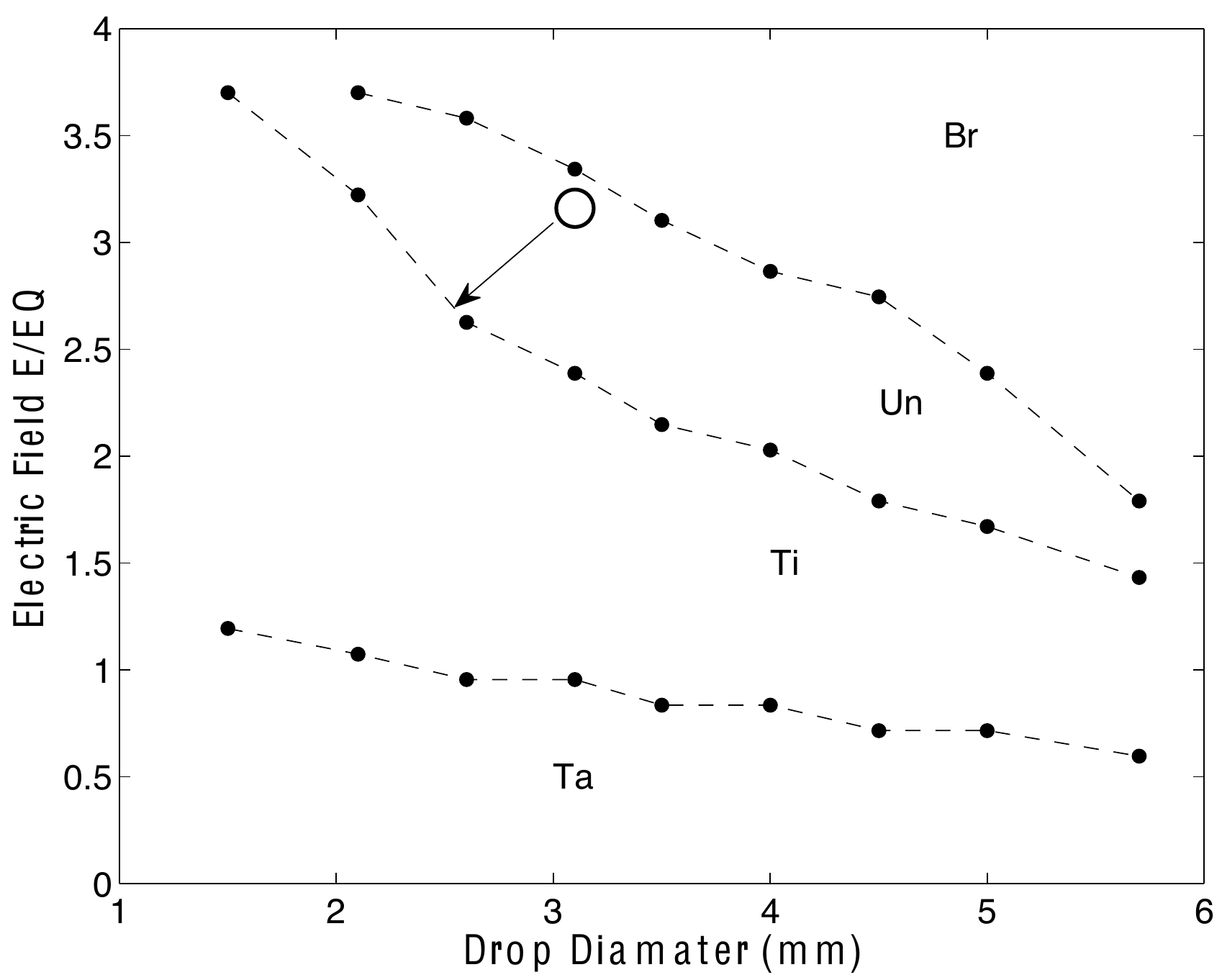}
  \put(10,27){\includegraphics[width=1.1in]%
      {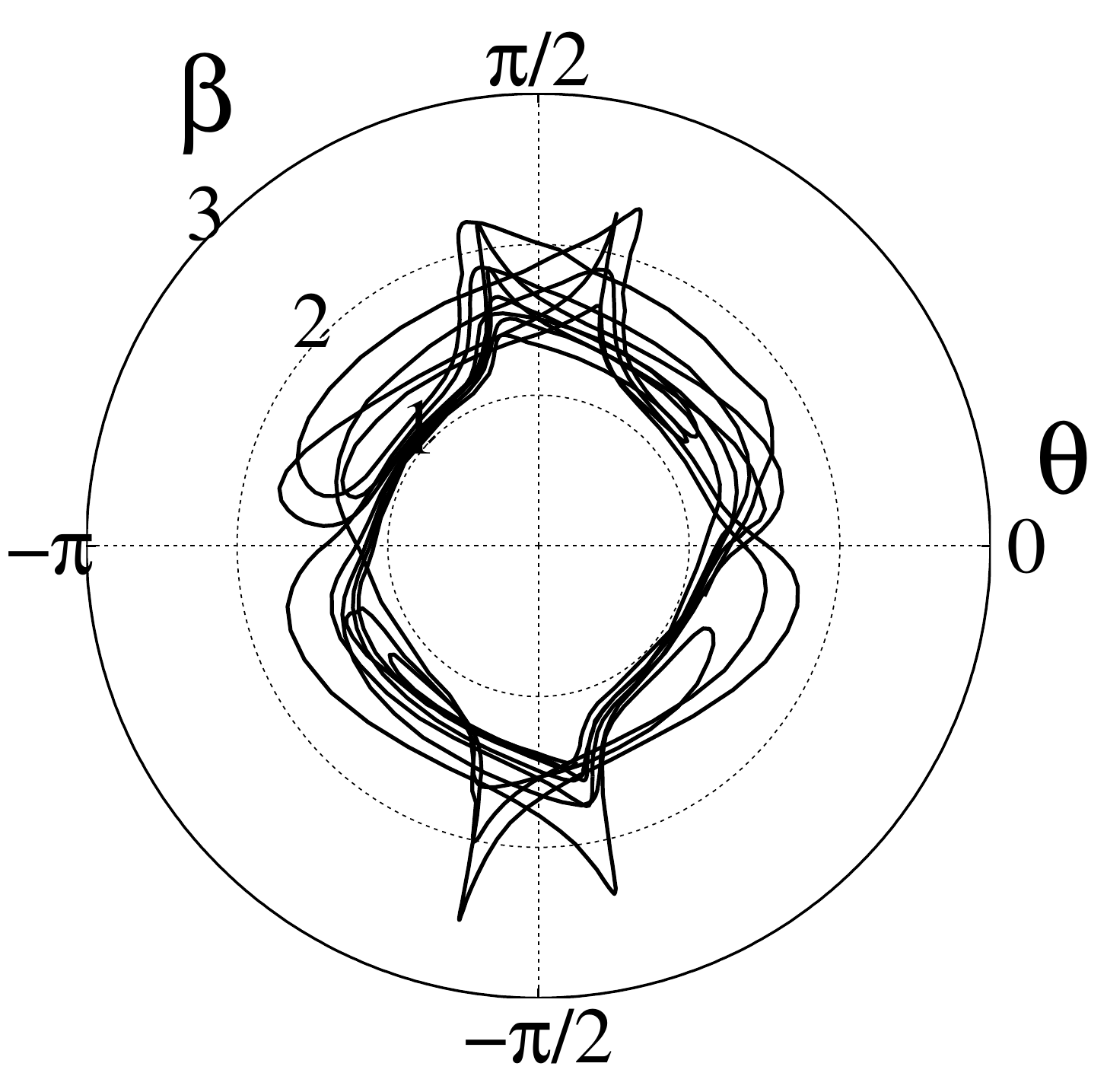}}
\end{overpic}

\begin{overpic}[width=3.15in]{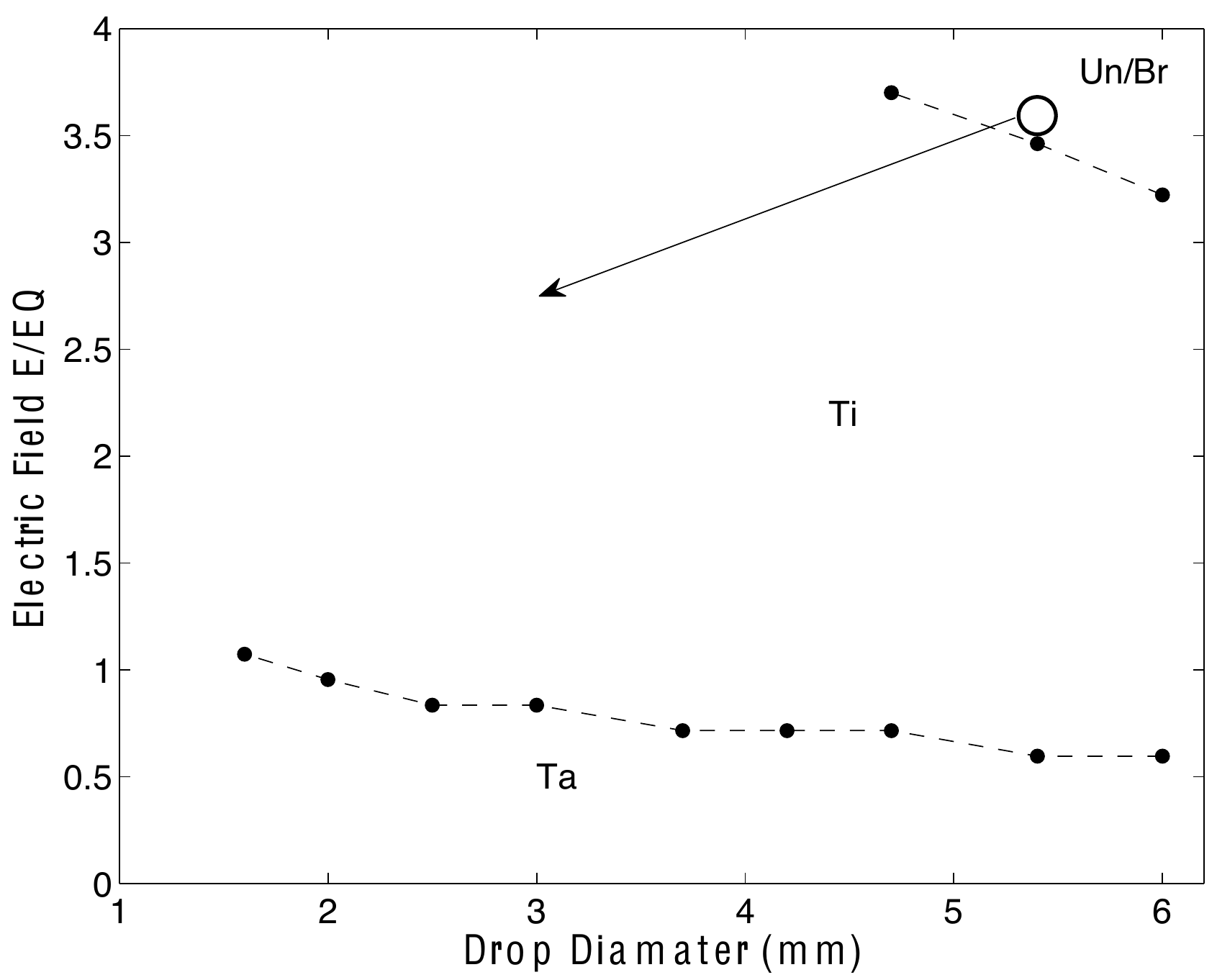}
  \put(10,30){\includegraphics[width=1.2in]%
      {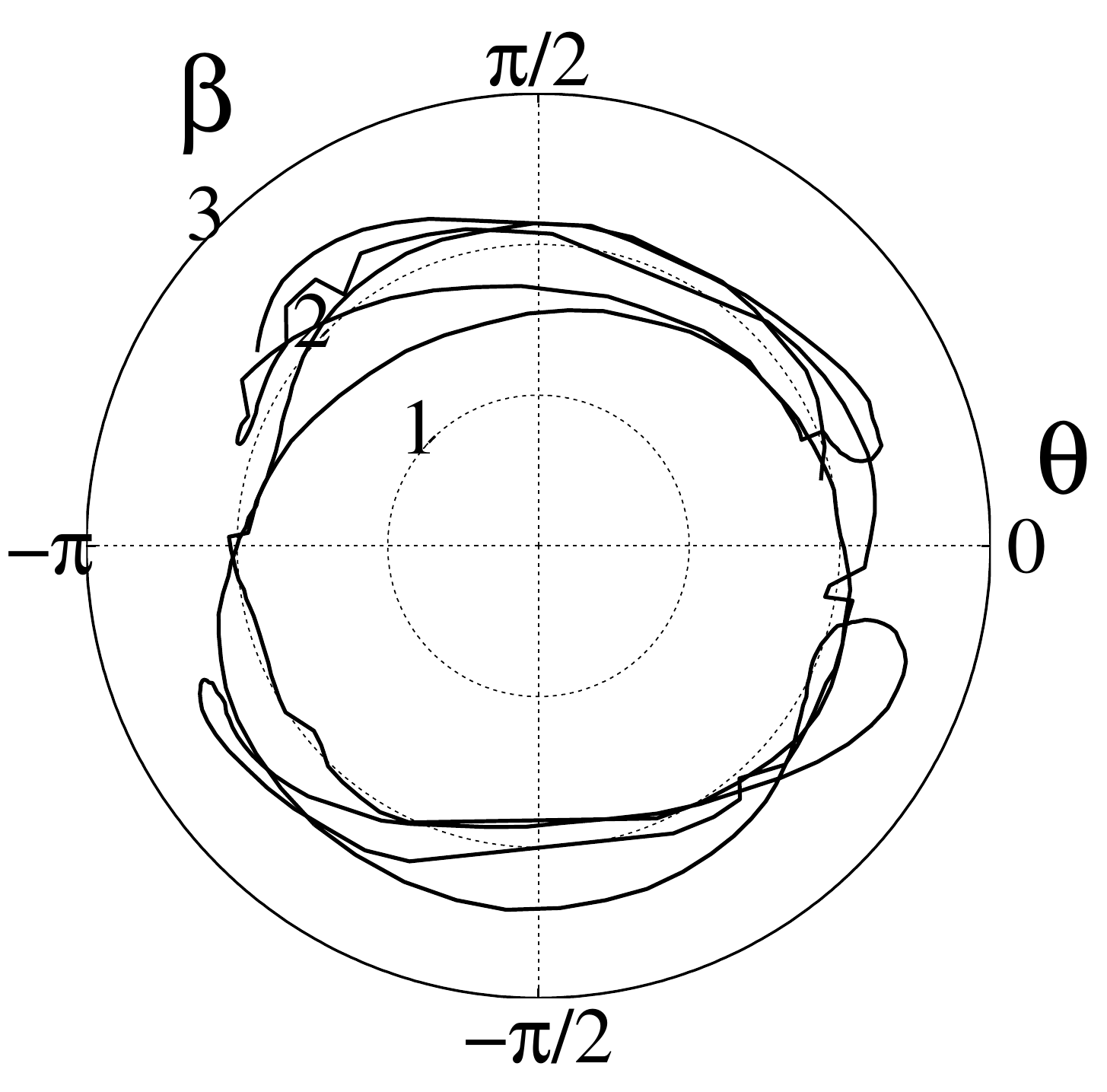}}
\end{overpic}
\caption{\footnotesize Phase diagrams for viscosity ratios $\lambda=$1, 4, 14. The electric field is slowly increased and steady-state behavior is observed.  Taylor (Ta) indicates axisymmetric flow and oblate deformation.  Tilted (Ti) indicates a tilted drop orientation with rotational flow. Unsteady (Un) indicates time dependent drop shape and orientation. Breakup (Br) indicates regions where drop breakup is observed.  The inset shows the time dependent behavior for one drop indicated on the chart with $\circ$, radial units of aspect ratio $\beta$ and angle units of drop orientation $\theta$. }
\label{fig6}
\end{figure}

The different unsteady behaviors are attributed to the interplay of shape and dipole dynamics. The more viscous the drop fluid, the higher resistance to fluid motion. Accordingly, high viscosity drop tend to behave as rigid particles and tumble in the electric field. They deform very little during the dipole rotation. In contrast, low viscosity drops undergo large deformations seen as shape oscillations. The limited deformation of high viscosity drops  allows for the fixed shape model  to accurately predict the transition to unsteady behavior.

In conclusion, in this Letter we report novel nonlinear dynamics of a droplet in uniform DC electric fields.
Droplet tumbling, shape oscillations, and chaotic rotations occur   under creeping  flow conditions, where nonlinear phenomena  are rare. We have constructed a phase digram of drop behaviors as a function of viscosity ratio and field strength. The experimental data qualitatively agrees with a theory which  models the drop as a fluid ellipsoid. The favorable comparison is surprising given the many simplifications in the theory and encouraging for future efforts to build a more refined and accurate model.
 The  nonlinear drop physics could 
 inspire  new approaches in electromanipulation for small-scale fluid and particle motion in microfluidic technologies.  
\bibliographystyle{unsrt}

\end{document}